\documentclass[prd, twocolumn, nofootinbib, tightenlines,preprintnumbers,notitlepage,longbibliography,superscriptaddress]{revtex4-1}

\usepackage{graphicx}
\usepackage{hhline}
\usepackage{amsmath}
\usepackage{amsfonts}
\usepackage[colorlinks=true,citecolor=blue,urlcolor=cyan,linkcolor=black]{hyperref}
\usepackage{tablefootnote}
\usepackage{multirow}
\usepackage{relsize}

\usepackage{aas_macros}

\newcommand{\ssum}{\mathsmaller{\sum}}

\usepackage{setspace}
\usepackage[hang]{footmisc} 
\usepackage{scalerel,stackengine}
\stackMath
\newcommand\what[1]{%
\savestack{\tmpbox}{\stretchto{%
  \scaleto{%
    \scalerel*[\widthof{\ensuremath{#1}}]{\kern-.6pt\bigwedge\kern-.6pt}%
    {\rule[-\textheight/2]{1ex}{\textheight}}
  }{\textheight}%
}{0.5ex}}%
\stackon[1pt]{#1}{\tmpbox}%
}

\usepackage{tikz} 
\usetikzlibrary{shapes,arrows,positioning,automata,backgrounds,calc,er,patterns}
\usepackage[compat=1.1.0]{tikz-feynman}

\definecolor{colorA}{HTML}{1E90FF}
\definecolor{colorB}{HTML}{228B22}
\definecolor{colorC}{HTML}{FF7F00}
\definecolor{colorD}{HTML}{4B0082}
\definecolor{colorE}{HTML}{B22222}

\definecolor{lgreen}{HTML}{32CD32}
\definecolor{lgray}{HTML}{D3D3D3}

\definecolor{teal}{HTML}{008080}
\definecolor{firebrick}{HTML}{B22222}
\definecolor{salmon}{HTML}{FA8072}
\definecolor{darkgreen}{HTML}{006400}

\definecolor{colortabpurple}{HTML}{9467bd}
\definecolor{colortaborange}{HTML}{ff7f0e}
\definecolor{colorpythongreen}{HTML}{008000}
\definecolor{colorpythongray}{HTML}{808080}
\definecolor{colortabblue}{HTML}{1f77b4}

\usepackage{graphicx}
\usepackage{dcolumn}
\usepackage{bm}
\usepackage[normalem]{ulem}
\usepackage{xcolor}

\graphicspath{ {figures/} }

\usepackage{epstopdf}
\newcommand{\be}{\begin{eqnarray}}

\newcommand{\ee}{\end{eqnarray}}

\newcommand{\bk}{\boldsymbol{k}}

\usepackage[export]{adjustbox}

\def\rhat{\hat{\boldsymbol{r}}}

\usepackage{suffix}
\usepackage{mathtools}

\DeclarePairedDelimiterX\MeijerM[3]{\lparen}{\rparen}%
{\begin{smallmatrix}#1 \\ #2\end{smallmatrix}\delimsize\vert\,#3}

\newcommand\MeijerG[8][]{%
  G^{\,#2,#3}_{#4,#5}\MeijerM[#1]{#6}{#7}{#8}}

\WithSuffix\newcommand\MeijerG*[7]{%
  G^{\,#1,#2}_{#3,#4}\MeijerM*{#5}{#6}{#7}}

\newcommand{\dd}{{\rm d}}

\begin{document}

\title{Neutrino Mass Constraints from kSZ Tomography}

\newcommand{\imperial}{Department of Physics, Imperial College London, Blackett Laboratory, Prince Consort Road, London SW7 2AZ, UK}

\newcommand{\jhu}{Department of Physics \& Astronomy, Johns Hopkins University, Baltimore, MD 21218, USA}

\newcommand{\perimeter}{Perimeter Institute for Theoretical Physics, 31 Caroline St N, Waterloo, ON N2L 2Y5, Canada}

\newcommand{\bengurion}{Department of Physics, Ben-Gurion University of the Negev, Be’er Sheva 84105, Israel}

\newcommand{\dartmouth}{Department of Physics and Astronomy, Dartmouth College, 6127 Wilder Laboratory, Hanover, NH 03755}

\newcommand{\uiuc}{Illinois Center for Advanced Studies of the Universe \& Department of Physics, University of Illinois at Urbana-Champaign, Urbana, IL 61801, USA}

\newcommand{\upenn}{Department of Physics and Astronomy, University of Pennsylvania,
209 South 33rd Street, Philadelphia, PA, USA 19104}

\author{Avery~J.~Tishue}
\email{atishue@illinois.edu}
\affiliation{\uiuc}
\affiliation{\dartmouth}
\affiliation{\jhu}

\author{Selim~C.~Hotinli}
\affiliation{\perimeter}

\author{Peter~Adshead}
\affiliation{\uiuc}
\affiliation{\upenn}

\author{Ely~D.~Kovetz}

\affiliation{\bengurion}

\author{Mathew~S.~Madhavacheril}
\affiliation{\upenn}

\date{\today}

\begin{abstract}

We forecast neutrino mass constraints using Stage IV CMB and large-scale structure surveys, focusing on kSZ tomography as an independent probe of the growth of cosmic structure. We take into account several realistic factors, including the kSZ optical depth degeneracy. Our baseline setup consists of CMB S4 temperature and polarization (but not lensing) information, DESI BAO, the LSST galaxy power spectrum, and a Planck like $\tau$ prior, yielding $\sigma(\sum m_\nu) = 32\, \rm{meV}$. Adding kSZ tomography improves this by a few percent, while a kSZ optical depth prior can push this improvement to over $15\%$, giving $\sigma(\sum m_\nu) = 27\, \rm{meV}$. When CMB lensing is included in the baseline setup, kSZ does not further improve neutrino mass constraints. We find promising prospects for a scenario combining futuristic CMB and galaxy surveys.

\end{abstract}

\maketitle

\section{Introduction\label{sec:intro}}

The determination of the total neutrino mass is among the most highly anticipated targets of forthcoming cosmological surveys~\citep{2009arXiv0912.0201L, Aghamousa:2016zmz, Abazajian:2016yjj, Ade:2018sbj, Abitbol:2019nhf}. Neutrinos, among the least understood elements of the Standard Model, play a critical role in both cosmology and astrophysics. Understanding their fundamental properties remains a central challenge in particle physics.  Precision measurements of neutrino flavor oscillations have accurately constrained the mass-squared differences between neutrino mass eigenstates, implying that the total mass of neutrinos must exceed $0.058\, \rm{eV}$ (see e.g. \cite{Capozzi:2017ipn, ParticleDataGroup:2018ovx}). This threshold represents a robust target within the standard cosmological model, offering a clear and testable objective for future observational campaigns. However, the small masses of neutrinos and their relativistic nature until late times lead to subtle effects on cosmological observables, making it challenging to isolate the neutrino signal from the effects of other parameters and uncertainties. Therefore, developing alternative methods to detect neutrino masses is not a straightforward endeavor that can be achieved without significant effort and innovation. Furthermore, the most precise cosmological constraints on the total neutrino mass, obtained through analyses of the cosmic microwave background (CMB) and baryon acoustic oscillations (BAO) data from the Dark Energy Spectroscopic Instrument (DESI)~\citep{DESI:2024mwx}, suggest a puzzling value below the lower bound established by neutrino flavor oscillation experiments~\citep{Craig:2024tky}, which may also be related to other anomalies in cosmology \cite{Loverde:2024nfi, Green:2024xbb}. The precise measurement of the sum of neutrino masses remains a significant, timely challenge, which motivates the integration of diverse and complementary data and methods ~\citep[e.g.][]{MoradinezhadDizgah:2021upg, Shmueli:2024npx, Racco:2024lbu}. Here, we explore the potential of a novel approach that combines measurements of the kinetic Sunyaev Zel'dovich (kSZ) effect~\citep{1980ARA&A..18..537S,1980MNRAS.190..413S,1970CoASP...2...66S} in the CMB with galaxy survey data.

\begin{figure*}[ht!]
    \includegraphics[width=1.6\columnwidth]{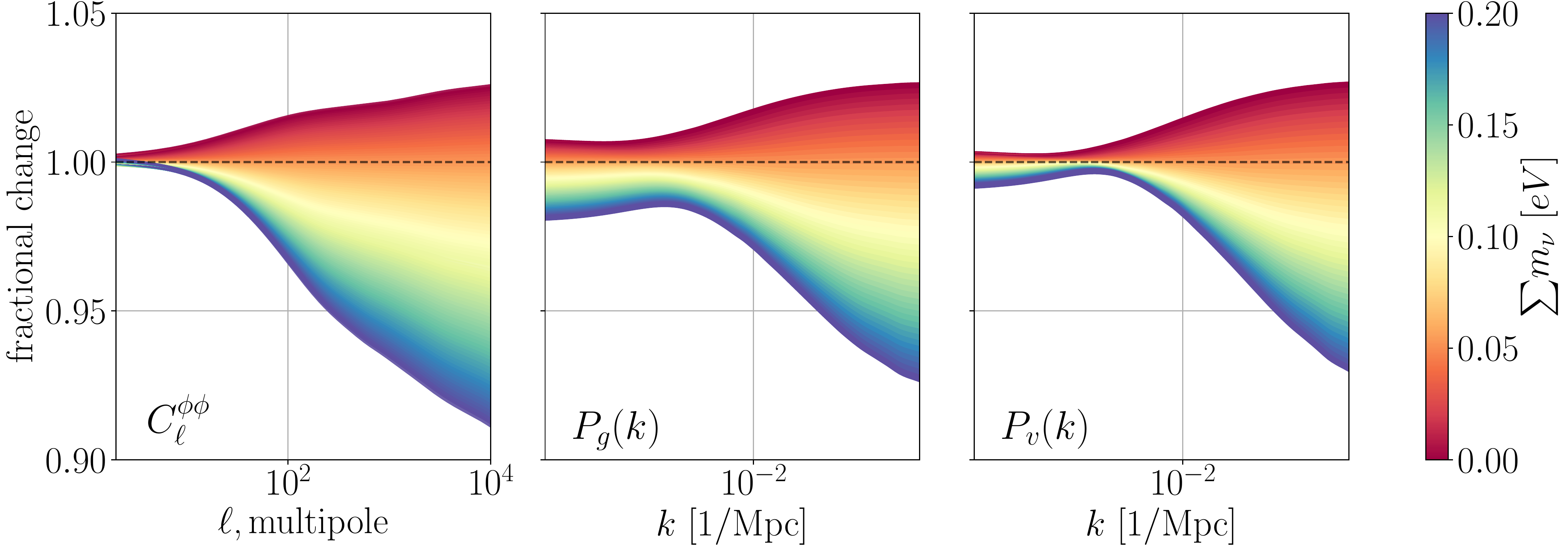}
    \vspace{-0.3cm}
    \caption{\textbf{Sensitivity of various power spectra to the sum of the neutrino masses at redshift $\boldsymbol{z=0.8}$}. The small-scale suppression of the lensing-potential power spectrum (left), the galaxy power spectrum (middle) and velocity power spectrum (right), due to massive neutrinos, is shown as a function of multipole $\ell$ and Fourier wave number $k$ . The blue end of the color spectrum corresponds to more massive neutrinos, $\ssum m_\nu= 0.2 \, \rm{eV}$ which further suppresses the various power spectra, while the red end of the spectrum correspond to massless neutrinos. The fractional changes in the spectra are normalized with respect to the normal hierarchy, $\ssum m_\nu = 0.06 \, \mathrm{eV}$ (black dashed line). } 
    \label{fig:sensitivityPk}
\end{figure*}

Measurement of the kSZ effect is poised to offer a potentially powerful new probe of cosmology in the coming decade: the next generation of cosmic microwave background (CMB) experiments such as the Simons Observatory (SO)~\citep{Ade:2018sbj}, Advanced SO (Adv. SO)~\citep{Abitbol:2019nhf} and CMB S4~\citep{Abazajian:2016yjj}, and galaxy surveys such as DESI~\citep{Aghamousa:2016zmz} and the Vera Rubin Observatory Legacy Survey of Space and Time (VRO LSST)~\citep{2009arXiv0912.0201L} will generate a wealth of new data with unprecedented precision on small scales. Interactions of CMB photons with the intervening large-scale structure induce correlations between CMB anisotropies and galaxy densities that carry valuable cosmological information~\citep[e.g.][]{Seljak:2008xr,Zhang:2015uta,Banerjee:2016suz,Schmittfull:2017ffw,Modi:2017wds,Deutsch:2017ybc,Cayuso:2019hen,Pan:2019dax,Hotinli:2019wdp,Hotinli:2020csk,Lee:2022udm,AnilKumar:2022flx,Hotinli:2022wbk,Cayuso:2021ljq,Ji:2021djj,Hotinli:2022jna,Kumar:2022bly,Hotinli:2023scz,Hotinli:2022jnt,Jain:2023jpy,SPT-3G:2024lko,CMB-S4:2024zgz}. In addition to the kSZ effect, these interactions include thermal SZ effect~\citep{1969Ap&SS...4..301Z,1970A&A.....5...84Z,1980ARA&A..18..537S,1972CoASP...4..173S,Sazonov:1999zp}, the integrated Sachs-Wolfe effect~\citep{1967ApJ...147...73S}, which include the moving lens effect~\citep{1983Natur.302..315B, 1986Natur.324..349G,Hotinli:2018yyc,2021PhRvD.103d3536H,Hotinli:2021hih,Hotinli:2024tjb,Hotinli:2024tjb,Beheshti:2024dxw}, and weak gravitational lensing (see e.g.~\citep{Lewis:2006fu} for a review). In particular, both the kSZ and moving lens effects are sourced due to peculiar motions of cosmological structure, making them useful probes of fundamental physics that affect the density field and the growth rate, a key example of which is massive neutrinos.

\begin{figure}[b!]
    \includegraphics[width=1.0\columnwidth]{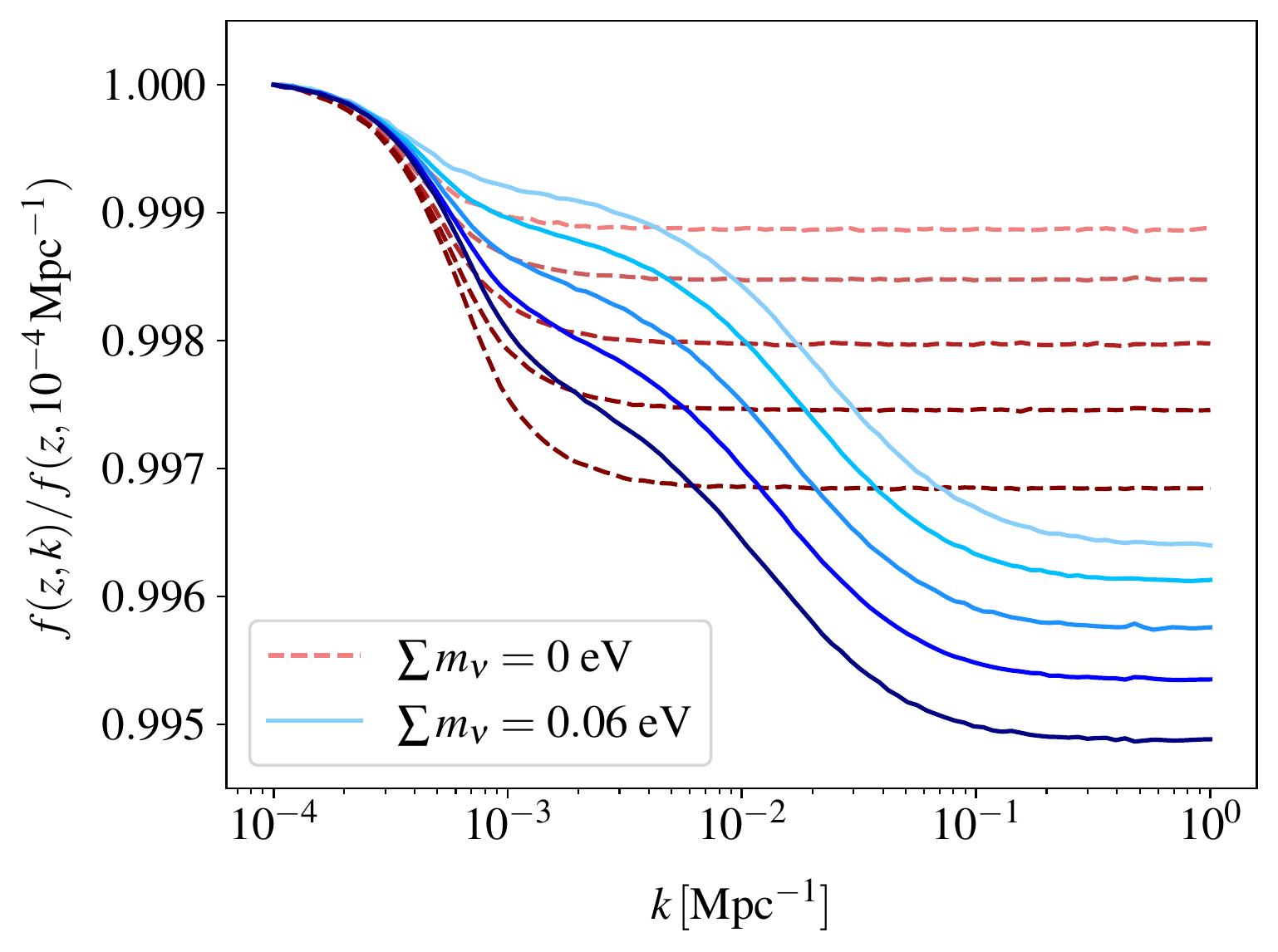}
    \vspace{-0.3cm}
    \caption{\textbf{Scale dependent growth rate $\boldsymbol{f(z,k)}$ due to massive neutrinos.} Massive neutrinos (blue, solid) alter scale dependent growth compared to massless neutrinos (red, dashed). We show five redshifts, $z \in [0.2,0.7,1.3,1.9,2.6]$, where darker lines indicate higher $z$. For each redshift, the curve is normalized to the growth at $k = 10^{-4} \, \mathrm{Mpc}^{-1}$. } 
    \label{fig:sensitivity_fofk}
     \vspace{-0.3cm}
\end{figure}

Massive neutrinos delay and suppress the growth of cosmological structure on scales below the neutrino sound horizon. This makes the cosmological growth rate and various cosmic observables sensitive to the sum of the neutrino masses $\sum m_\nu$. The sensitivity of cosmological fluctuations to massive neutrinos can be seen in Figs.~\ref{fig:sensitivityPk}~and~\ref{fig:sensitivity_fofk}, where we show the fractional change in the lensing potential and the galaxy and velocity power spectra and growth factor $f(k)$ due to the varying neutrino mass compared to the fiducial value of the normal hierarchy $\sum m_\nu=0.06~\mathrm{eV}$.  The effect on these power spectra reaches $1$-$10 \%$ on small scales $k\sim10^{-1}\,{\rm Mpc}^{-1}$ ($\ell \sim 10^4$). Therefore, these observables provide important probes of $\sum m_\nu$. Specifically, joint analysis of large-scale structure surveys with the kSZ signature in the CMB makes tomographic reconstruction of radial cosmological velocities possible~\citep{Deutsch:2017cja, Smith:2018bpn,Hotinli:2021hih}, in principle providing a new way of measuring $\sum m_\nu$ that is complementary to other probes such as weak gravitational lensing of the CMB and galaxies~\citep[e.g.][]{PhysRevLett.91.041301,TopicalConvenersKNAbazajianJECarlstromATLee:2013bxd}, measurements of galaxy-survey redshift-space distortions (RSDs) and galaxy clustering~\citep[e.g.][]{LoVerde:2016ahu,Upadhye:2017hdl}, or line intensity mapping \cite{Shmueli:2024npx}.

In this paper, we investigate the power of kSZ tomography for constraining the sum of the neutrino masses. Indeed, kSZ velocity reconstruction has emerged as a promising probe of fundamental physics, for example to probe primordial non-Gaussianity \cite{Munchmeyer:2018eey, Adshead:2024paa, Lague:2024czc}, dark energy and modifications to gravity \cite{DeDeo:2005yr, Bhattacharya:2007sk, Mueller:2014nsa, Pan:2019dax}, and as we focus on here, the imprint of massive neutrinos \cite{Mueller:2014dba}. In Ref. \cite{Mueller:2014dba}, it was suggested that the pairwise kSZ effect with Stage IV cosmological surveys could improve neutrino mass constraints compared to temperature, polarization, and lensing information from the Planck satellite \cite{Aghanim:2018eyx}. Here, we instead focus on kSZ tomography with the long-wavelength velocity reconstruction method \cite{Deutsch:2017ybc, Smith:2018bpn}, and provide an up-to-date forecast on the role of kSZ measurements for cosmological neutrino mass inference. 

Our primary objective is to evaluate the additional information kSZ tomography provides beyond that offered by the galaxy and CMB surveys that enable the velocity reconstruction, rather than to compare Stage IV kSZ tomography to earlier CMB bounds. In particular, we focus on the role of kSZ tomography as an independent probe of the growth of cosmic structure, excluding CMB lensing and explicit growth information from the galaxy power spectrum in our baseline forecasts. Our forecasts carefully account for realistic factors, including uncertainties in galaxy bias and the cosmological model, photometric redshift errors, CMB foregrounds, and contributions from independent probes such as BAO and CMB lensing. We fully account for the kSZ optical depth degeneracy by marginalizing over a velocity reconstruction bias. Consistently, we find that within the context of Stage IV CMB and LSS surveys, kSZ tomography contributes limited additional information beyond the galaxy clustering and CMB data already integral to velocity reconstruction. 

Some studies have questioned the extent to which linear-scale observations can provide additional information about massive neutrinos beyond what is already encoded in the dark matter density field and redshift-space distortions~\cite{Bayer:2021kwg}. There are several reasons one may expect kSZ tomography to improve neutrino mass inference. On linear scales, the velocity mode is proportional to the product of the density field and the cosmic growth rate, $f$, upon which the neutrinos imprint a distinct scale- and redshift-dependent signature, see Fig.~\ref{fig:sensitivity_fofk}. While the effect is small, it is possible in principle that this serves as an additional source of information about the neutrino masses that velocities can provide. Furthermore, velocities, like galaxies, trace the underlying matter power spectrum, but with a distinct bias, offering the potential to mitigate large-scale sample variance that challenges single-tracer analyses. This sample variance cancellation can help disentangle cosmological parameters from astrophysical bias parameters, thereby enhancing parameter constraints. Finally, by providing an additional measurement of the matter power spectrum on the largest scales, kSZ tomography can act as an anchor by pinning down the amplitude and shape of the matter power spectrum on large scales, thereby improving the sensitivity of the galaxy power spectrum on smaller scales where the impact of massive neutrinos is more prominent. In this way, one may imagine that kSZ tomography can aid in the measurement of the imprint of massive neutrinos despite not having great sensitivity to those scales on its own. However, as we demonstrate in this paper, the potential of kSZ tomography to significantly enhance constraints on massive neutrinos in Stage IV cosmological surveys is limited. Of the above possibilities, we find that the majority of the additional neutrino mass information that kSZ provides is from the measurement of scale-dependent growth. However, information from the CMB, even without lensing, is so strongly constraining that it effectively overshadows this contribution. Alternatively, if the kSZ optical depth degeneracy can be circumvented, kSZ tomography may appreciably improve neutrino mass constraints, even if artificially neglecting scale-dependence in $f(k)$, but in practice this contribution is small when CMB lensing is included. Ultimately, this leaves kSZ tomography with a complementary but less distinctive role.

The remainder of this paper is organized as follows. In Sec.~\ref{sec:vel_rec} we briefly review kSZ and galaxy tomography. Details regarding our forecasts and methods are outlined in Sec.~\ref{sec:forecasts}. Our results on neutrino mass constraints and subsequent analysis, including the impact of bias modeling, the kSZ optical depth degeneracy, parameter degeneracies, model extensions, and varying experimental configurations, are presented in Sec.~\ref{sec:results}. We discuss the implications of this work and prospects for future progress in Sec.~\ref{sec:discussion}.

\section{KSZ and galaxy tomography}\label{sec:vel_rec}

KSZ tomography is the extraction of cosmological information from the kSZ effect in the CMB together with a tracer of the density field, such as a galaxy survey, by reconstructing the radial velocity field. Following Ref.~\cite{Smith:2018bpn}, the kSZ effect can be described in a ``snapshot" geometry, where the Universe is taken to be a 3D box at some time $t$ and redshift $z$. The CMB temperature anisotropy induced by the kSZ effect due can then be written as 
\be
\Theta_{\rm kSZ}(\boldsymbol{\theta})=K(z)\int_0^R\dd r\, q_{r}(\boldsymbol{x}), 
\ee
where $\boldsymbol{x}\equiv\chi(z)\boldsymbol{\theta}+r\hat{\boldsymbol{r}}$, $\rhat$ is the unit vector in the radial direction, $\chi(z)$ is the comoving distance, $\Theta(\boldsymbol{\theta})=\Delta T(\boldsymbol{\theta})/\bar{T}$ is the CMB temperature fluctuation divided by the sky-averaged CMB temperature $\bar{T}$, $\boldsymbol{\theta}$ is the angular direction on the sky, and $K(z)$ is the kSZ radial weight function in units of $1/{\rm Mpc}$ \cite{Smith:2018bpn}. {The electron-momentum field, projected onto the radial direction, is $q_{r}(\boldsymbol{x})=\delta_e(\boldsymbol{x})v_{r}(\boldsymbol{x})$.} 

While the small-scale free electron density fluctuation is determined mainly by electron astrophysics, the bulk radial velocity field has been shown to contain significant cosmological information (see e.g. Refs.~\cite{Smith:2018bpn,Munchmeyer:2018eey}). A quadratic estimator for the large-scale radial velocity field $\hat{v}_r$ can be constructed using a density tracer, e.g. galaxies, and the CMB temperature field \cite{Smith:2018bpn}. The reconstruction noise of this radial velocity estimator can be shown to satisfy 
\begin{align}
\label{eq:ksz_tomo}
\frac{1}{N_{\hat{v}_r \hat{v}_r}(\boldsymbol{k}_L, \mu, z)}=&\frac{K(z)^2}{\chi^2(z)} \\
& \times \int \frac{k_s\dd k_s}{2\pi}\frac{P^2_{ge}(k_s)}{P^{\rm obs}_{gg}(k_s, k_L, \mu,z)C_{\ell = k_s \chi(z)}^{TT,\mathrm{\,obs}}}, \nonumber
\end{align} 
where $\mu = \hat{k}\cdot \hat{r}$, $\bk$ is the three-dimensional Fourier wavevector and the integral is over small-scale Fourier modes $k_S$ that contribute to reconstruction of the large scale modes $k_L$. The range of $k_S$ for the velocity reconstruction depends on the survey details, and for the surveys we consider here, the integral is dominated by the range  $k_S \in [0.1,10] \,\mathrm{Mpc}^{-1}$. The spectra in the denominator of the integrand in Eq.~\eqref{eq:ksz_tomo} correspond to the observed power spectra of galaxies, $P^{\rm obs}_{gg}$, and the CMB temperature, $C_\ell^{TT,\rm \, obs}$, while $P_{ge}$ is the galaxy-electron cross spectrum. The observed galaxy power spectrum is given by 
\begin{align}
P_{gg}^{\mathrm{obs}}(k_S,k_L,\mu,z) = P_{gg}(k_S,z) + N_{gg}(k_L,\mu,z), 
\end{align}
where the noise $N_{gg}$ contains both the shot noise, determined by the galaxy number density $n_g$, and photometric redshift errors, given by 
\be
W(k,\mu,z)=\exp \left(-\mu^2k_L^2\sigma_z^2/2H^2(z)\right),\label{eq:Wphotoz}
\ee 
such that $N_{gg} = 1/n_g W^2$. Here, $H(z)$ is the Hubble parameter, and $\sigma_z$ is the photo-$z$ error. 

The reconstructed velocity field can serve as a tracer of the underlying matter power spectrum and is sensitive to the growth of cosmic structure. On large scales, linear perturbation theory relates the velocity mode to the density mode via the continuity equation,
\begin{align}
   v(\bm{k}) = i(f(k,a)aH/k)\delta_m (\bm{k}),
\end{align}
where the growth rate $f(k,a)$ is related to the growth function $D(k,a) = (P_{mm}(k,a)/P_{mm}(k,a=1))^{1/2}$ via $f = d\ln D/ d\ln a$. The reconstructed radial velocity mode $\hat{v}_r$ is related to the true velocity mode $v$ via 
\begin{align}
    \hat{v}_r(\bm{k}) = b_v\mu v(\bm{k}),
\end{align}
where $b_v$ is the velocity reconstruction bias, which accounts for the potential mismodeling of the true galaxy-electron cross spectrum  $P_{ge}^{\rm{true}}$. It is given by \cite{Smith:2018bpn}
\begin{align}
b_v &= \frac{\int dk_S F(k_S) P^{\mathrm{true}}_{ge}(k_S)}{\int dk_S  F(k_S) P^{\mathrm{fid}}_{ge}(k_S)} \label{eq:bv_def},
\end{align}
where 
\begin{align}
F(k_S) &= k_S \frac{P^{\mathrm{fid}}_{g e}(k_S)}{P^{\mathrm{obs}}_{g g}(k_S) C_{\ell = k_S \chi}^{TT, \mathrm{obs}} }.
\end{align}
Here $P_{ge}^{\rm fid}$ is the fiducial choice for modeling the galaxy-electron cross correlation in the velocity reconstruction. The imperfect knowledge of $P_{ge}$, manifesting in the form of this parameter $b_v$, is known as the kSZ optical depth degeneracy. To address astrophysical uncertainties in modeling the galaxy-electron cross-correlation, we marginalize over $b_v$, introducing a potentially important degeneracy that can reduce the cosmological constraining power of kSZ tomography. 

From the galaxy survey and kSZ velocity reconstruction, we can construct the power and cross-spectra, which are related to the matter power spectrum $P_{mm}$ via
\begin{align}
P_{gg}(k,\mu,z) &= b_g^2(k,\mu,z) P_{mm}(k,z) \label{eq:Pgg} \, , \\
P_{g\hat{v}_r}(k,\mu,z) &=b_g(k,\mu,z) b_v(z) \mu \left(\frac{faH}{k}\right)P_{mm}(k,z)\,,\\
P_{\hat{v}_r \hat{v}_r}(k,\mu,z) &= b_v^2(z) \mu^2 \left(\frac{faH}{k}\right)^2P_{mm}(k,z) \label{eq:Pvrvr} \,,
\end{align}
where we take as our model for the galaxy bias
\begin{align}
    b_g(k, \mu, z) = b_1(z)+b_{\rm rsd}(z)f\mu^2+b_2(z)k^2. \label{eq:gal_bias}
\end{align}
Here, we have the linear bias $b_1(z)$, the $f\mu^2$ Kaiser effect redshift-space distortion (RSD) term \cite{Kaiser:1987qv} with a bias $b_{\rm rsd}(z)$ that arises from anisotropic selection effects \cite{Obuljen:2020ypy}, and the lowest order gradient bias $b_2(z)$ which parameterizes non-linear, small-scale clustering. In addition to the kSZ effect, the transverse velocity field can be reconstructed using the moving lens effect, providing another differently biased tracer of the matter power spectrum. However, our analysis found that including transverse velocities from the moving lens effect resulted in negligible impact due to its significantly lower signal-to-noise ratio compared to kSZ tomography. Consequently, we do not include the moving lens effect in our study.

The galaxy power spectrum is a key cosmological probe, with the BAO signature offering robust distance measurements and sensitivity to various cosmological parameters through its phase and amplitude ~\citep[see e.g.][]{2010MNRAS.401.2148P}. Beyond BAO, the broadband power spectrum constrains $H(z)$, $D_A(z)$, the matter-radiation equality peak, Silk damping scale, and RSD effects, which are tied to the growth rate $f$. We assess the constraining power of both BAO and the full galaxy power spectrum at $k \lesssim 0.1 \, \text{Mpc}^{-1}$ for $\sum m_\nu$. 

\section{Forecast Details}\label{sec:forecasts}
In our forecasts, we consider a combination of galaxy and kSZ tomography, BAO, and CMB temperature, polarization, and lensing information. For the galaxy power spectrum, we take specifications chosen to match the anticipated redshift range and volume of LSST~\citep{2012JCAP...04..034H,LSSTDarkEnergyScience:2018jkl,Munchmeyer:2018eey}. We consider five redshift boxes with redshifts, volumes, galaxy number densities, and linear galaxy biases in Table \ref{tab:VRO_specs}.

\begin{table}[h!]
    \centering
    \begin{tabular}{|c|c|c|c|}
        \hline
        $z$ & $V\,[\rm{Gpc}^{3}]$ & $n_g\, [ \rm{Mpc}^{-3} ]$ & $b_1$ \\
        \hhline{|=|=|=|=|}
        0.2 & 5.2 & $5\times 10^{-2}$ & 1.05 \\ 
        0.7 & 43.6 & $2\times 10^{-2}$ & 1.37 \\
        1.3 & 75.9 & $6\times 10^{-3}$ & 1.79 \\
        1.9 & 89.3 & $1.5\times 10^{-3}$ & 2.22 \\
        2.6 & 119.9 & $3\times 10^{-4}$ & 2.74 \\
        \hline
    \end{tabular}
    \caption{Redshift, volume, galaxy density, and linear galaxy bias for each of the five redshift bins used in our forecasts.}
    \label{tab:VRO_specs}
\end{table}

To compute the kSZ velocity reconstruction noise, following Ref.~\cite{Hotinli:2021hih}, the total CMB power spectrum gets contributions from weak gravitational lensing, the moving lens effect, kSZ (both from reionization and late times), and the experimental noise. We take five frequency bins between $\sim38-280\rm GHz$ with white noise rms $\Delta_T$ and survey beam full-width at half maximum $\theta_{\rm FWHM}$ as defined in Ref.~\citep{Hotinli:2021hih} that match various upcoming CMB experiments. We define the white noise spectrum at each multipole as
\be
N_\ell=\Delta_T^2\exp\left[\frac{\ell(\ell+1)\theta^2_{\rm FWHM}}{8\ln2}\right]\,,
\ee
and consider frequency-dependent thermal Sunyaev Zel'dovich and cosmic infrared background foregrounds, as well as radio point sources, and perform the standard internal linear combination (ILC) cleaning to calculate the total CMB variance on the measured blackbody signal.

We define a diagonal noise matrix for the tomographic galaxy density and reconstructed-velocity fields as~\citep{Hotinli:2021hih}
\be
N_{ij}(k,\mu,z) = \mathrm{diag} [N_{\hat{v}_r \hat{v}_r},N_{gg} ], \label{eq:Nij}
\ee
where we have suppressed the $k$, $\mu$, and $z$-dependence on the right hand side. Similarly a signal matrix can be defined as
\be
S_{ij} = \begin{bmatrix}
P_{\hat{v}_r \hat{v}_r} & P_{\hat{v}_r g}  \\
P_{\hat{v}_r g}  & P_{gg}  
\end{bmatrix}. \label{eq:Sij}
\ee
For each redshift bin, an ensemble Fisher information matrix can be computed from these observables, 
\be
F_{ab} = \frac{V}{2} \int^\infty_0 \frac{k^2 dk}{4\pi^2} \int^1_{-1} d\mu \, \mathrm{Tr} \left[C^{-1} \frac{\partial S}{\partial \pi_a} C^{-1} \frac{\partial S}{\partial \pi_b} \right],
\ee
where the $(a,b)$ indices run over the free parameters $\pi$ in the forecast. Here the total covariance is  $C = S + N$  (with the $(k,\mu,z)$ dependence suppressed for brevity), and $V$ is the survey volume for each redshift bin. The trace is over the (suppressed) observables indices $(i,j)$ appearing in Eqs.~(\ref{eq:Nij}) and (\ref{eq:Sij}), and the integral over Fourier wavenumber $k$ is over large scale modes $k_L$ where the galaxy and velocity spectra are measured. In practice this range is bounded below\ by the volume of the snapshot geometry box, $k_L>k_{\mathrm{min}} = \pi/V^{1/3}$, and bounded above by the smallest $k_S$ used in the reconstruction, which we take to be $k_L < k_{\mathrm{max}} = 0.1 \, \mathrm{Mpc}^{-1}$.

We calculate a similar information matrix for the CMB temperature, polarization, and lensing reconstruction spectra using the publicly available codes \texttt{Fisherlens}\footnote{\url{https://github.com/ctrendafilova/FisherLens}} and \texttt{class\_delens}\footnote{\url{https://github.com/selimhotinli/class_delens}}~\cite{Hotinli:2021umk}, which account for non-Gaussian covariances due to lensing. We use the public \texttt{pyfisher}\footnote{\url{https://github.com/msyriac/pyfisher}} code for DESI BAO.

\section{Results \label{sec:results}}

\subsection{Baseline Forecasts}\label{sec:forecast_baseline}
We first present the results of our baseline forecast. We assume a CMB S4-like survey with a $1.4'$ beam and $\Delta_T = 1.4 \, \mu K'$. In our information matrix, we marginalize over $\sum m_\nu$ with a fiducial value of $0.06 \, \rm{}eV$, six $\Lambda$CDM parameters $\{ \Omega_b\, h^2,\,\Omega_{\rm cdm}h^2,\,A_s,\,n_s,\,\tau,\,H_0\}$ with fiducial values set to $\{0.022,0.12,{2.2\!\times\!10^{-9}},0.965,0.06,67.7\,{\rm km}/{\rm s}/{\rm Mpc}\}$, as well as the three galaxy bias parameters in Eq.~(\ref{eq:gal_bias}), and the velocity reconstruction bias $b_v(z)$. The galaxy and velocity reconstruction bias parameters are defined independently in each redshift bin. The fiducial values for the linear bias are given in Table \ref{tab:VRO_specs}. For the remainder of the bias terms, we set the fiducial values $b_{\rm rsd}(z)=1$, $b_v(z)=1$, and $b_2(z)=0$.

We are interested in both the raw power of future data sets for constraining $\sum m_\nu$, as well as the relative change in this constraining power as the experimental configuration varies. We evaluate this relative change by dividing the $1 \sigma$ measurement uncertainty of each configuration by the uncertainty from a reference configuration, which is denoted in brackets in the upper axis of our figures.
\begin{figure}[b!]
    \centering
\includegraphics[width=1.\columnwidth]{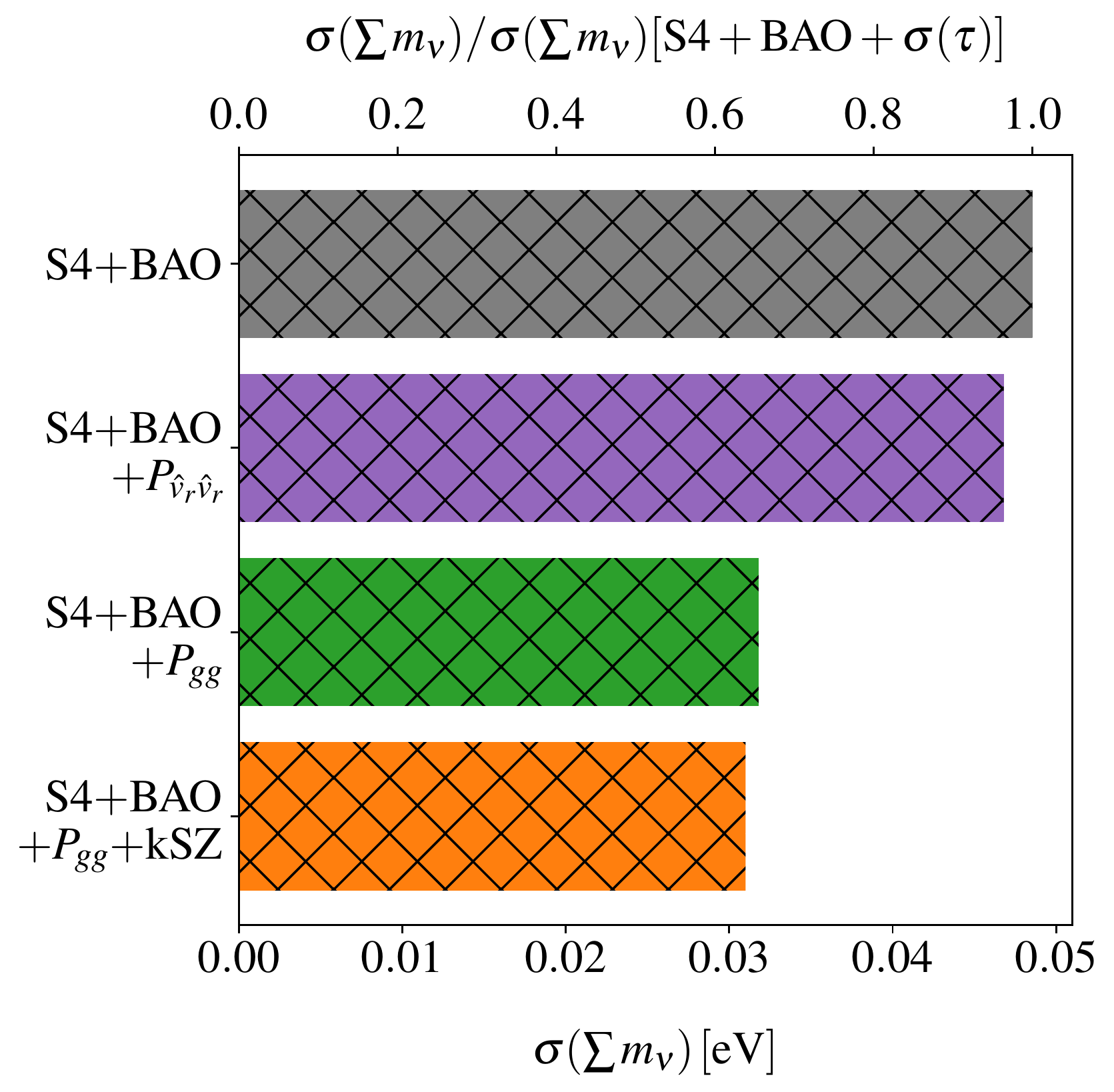}
    \caption{Neutrino mass constraints from various experimental configurations in our baseline forecasts with a Planck-like $\tau$ prior of $\sigma(\tau) = 0.0075$. The lower axis indicates the forecasted 1$\sigma$ error bar in $\rm{eV}$. The upper axis divides the $1\sigma$ error bar from each configuration by that from S4$+$DESI BAO$+\sigma(\tau)$, showing the fractional change in the constraining power. As described in the main text, all bars exclude CMB lensing information and marginalize over $b_{\rm{rsd}}$, such that kSZ-reconstructed velocities are the only direct probe of cosmic growth. The purple bar adds only information from the reconstructed radial velocity power spectrum, while the orange bar labeled ``$+\rm{kSZ}$" adds both $P_{g\hat{v}_r}$ and $P_{\hat{v}_r\hat{v}_r}$.}
    \label{fig:mnu_baseline}
\end{figure}

Our primary focus is to investigate the constraining power of kSZ tomography as an independent probe of the growth of cosmic structure for the purpose of neutrino mass constraints. To do so, we do not include any information from CMB lensing in our baseline forecast: we use unlensed $T$ and $E$ CMB spectra for the CMB Fisher matrices, and do not include the lensing reconstruction power spectrum $C_{\ell}^{\kappa \kappa}$ in the observables. This, in tandem with marginalizing over $b_{\rm{rsd}}$ that multiplies $f\mu^2$ in the galaxy bias, ensures that neither the CMB nor the galaxy survey on their own provide direct information about cosmological growth. Later, in Sec.~\ref{sec:lensing}, we return to the role of CMB lensing. Figure~\ref{fig:mnu_baseline} shows the results of our baseline forecast, where we consider neutrino mass constraints furnished by combinations of CMB temperature and $E$ mode polarization, DESI BAO, galaxy and kSZ tomography, and a Planck-like $\tau$ prior of $\sigma (\tau)=0.0075$. We note that while we elect to include a $\tau$ prior for our baseline results, it has little impact when all CMB lensing information is excluded. We comment on the role of the $\tau$ prior in Sec.~\ref{sec:lensing}. We take as a reference configuration the combination of CMB S4 and DESI BAO, in conjunction with the $\tau$ prior, which we label S4$+$BAO$+\sigma (\tau)$. We find that this reference configuration will achieve $\sigma(\sum m_\nu) \approx 0.0486 \,\rm{eV}$, enabling slightly over one-sigma evidence for the normal mass hierarchy. Compared to the reference configuration, the addition of the galaxy power spectrum reduces the measurement error by approximately $35\%$ to $\sigma(\sum m_\nu) \approx 0.0318 \rm{eV}$. However, kSZ tomography offers limited constraining power. First, the addition of only the radial velocity power spectrum to the reference configuration brings the measurement error to $\approx 0.0468 \, \rm{eV}$, only about $3.5\%$ better than S4+DESI BAO+$\sigma(\tau)$. Likewise, compared to S4+DESI BAO+$\sigma(\tau)+P_{gg}$, the addition of kSZ tomography, i.e. both the radial velocity power spectrum and the radial velocity-galaxy cross-spectrum, offers marginal additional constraining power, yielding $\sigma(\sum m_\nu) \approx 0.0310 \, \rm{eV}$, approximately $2.5\%$ tighter. If information from BAO is excluded, the constraints on the neutrino masses degrade significantly, but importantly the role of kSZ tomography remains limited,  offering a $ \approx 4\%$ improvement compared to S4+$\sigma(\tau) + P_{gg}$.

\begin{figure}[b!]
    \centering
\includegraphics[width=1.0\columnwidth]{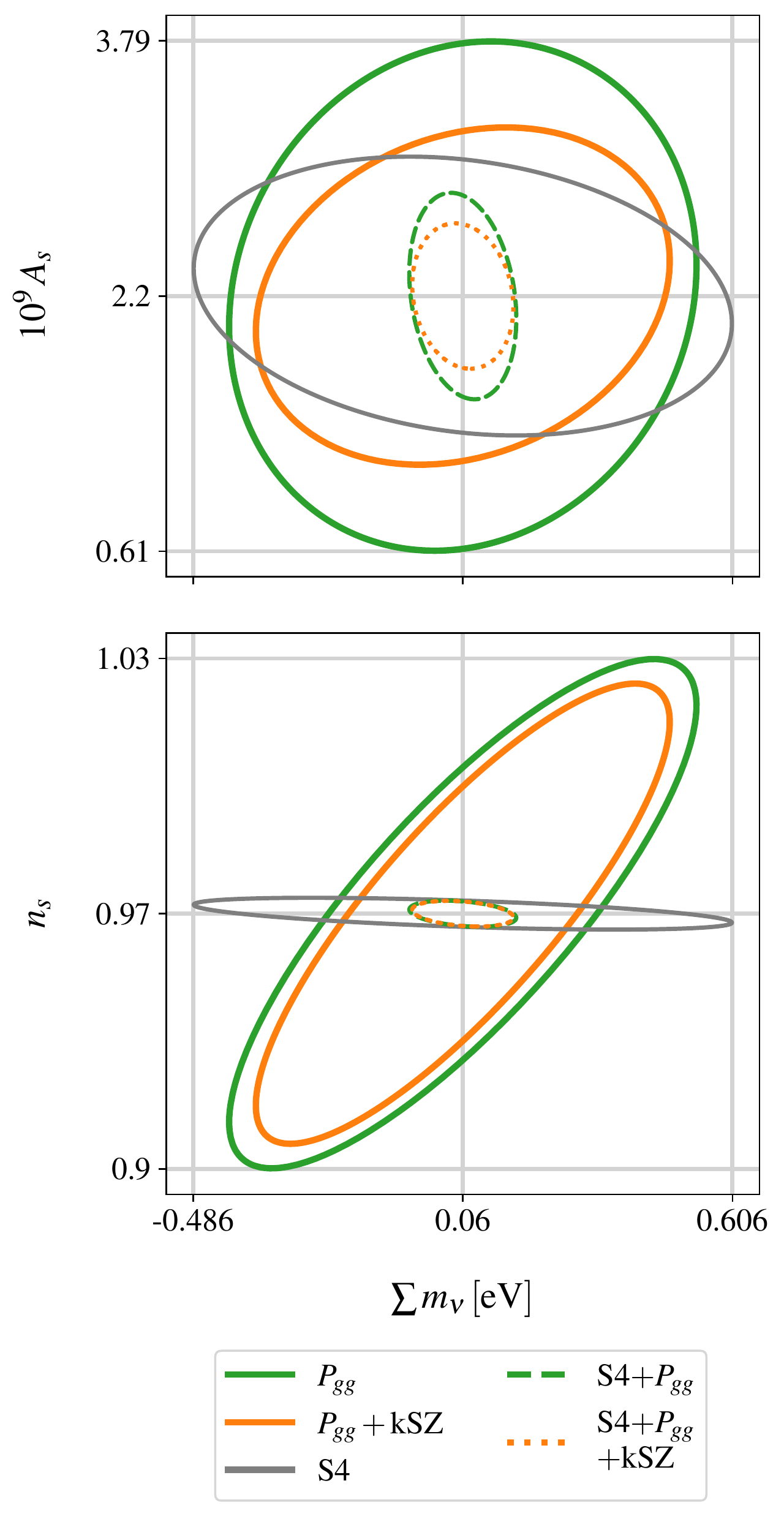}
    \caption{Forecasts for $1\sigma$ uncertainty contours in the $(A_s, \sum m_\nu)$ (top) and $(n_s, \sum m_\nu)$ (bottom) planes for CMB S4, galaxy, and kSZ tomography, and combinations thereof. The top and bottom panels share a horizontal axis. The contours contain no $\tau$ prior and no information from BAO or CMB lensing.  In the lower panel the dashed green (S4$+P_{gg}$) and orange (S4$+P_{gg}+\rm{kSZ}$) curves lie nearly on top of each other.}
    \label{fig:mnu_As_contours}
\end{figure}

To distill the above results, we show in Fig.~\ref{fig:mnu_As_contours} the uncertainty contours of $1\sigma$ in the $(A_s, \sum m_\nu)$ and $(n_s, \sum m_\nu)$ planes for several demonstrative experimental setups. The solid green and orange contours correspond, respectively, to galaxy tomography alone and the combination of galaxy and kSZ tomography, with no information from the CMB or BAO. Here, galaxy tomography alone gives $\sigma(\sum m_\nu)=0.311 \, \rm{eV}$, and the addition of kSZ tomography, i.e. $P_{g\hat{v}_r}$ and $P_{\hat{v}_r \hat{v}_r}$, gives $\sigma(\sum m_\nu)=0.276\,\rm{eV}$, an approximately $10\%$ improvement. Importantly, if ignoring scale-dependent growth, i.e. artificially enforcing $f=f(z)$ only, the improvement from kSZ becomes only about $1 \%$.  This indicates that in principle, the addition of kSZ tomography allows for better measurement of massive neutrinos compared to the galaxy power spectrum specifically by acting as an independent probe of scale-dependent growth. However, as Fig.~\ref{fig:mnu_baseline} indicates, the kSZ-driven improvement in neutrino mass constraints nearly disappears once other probes are included in the analysis, e.g. S4+DESI BAO. In fact, the essential picture is even simpler. To show this, we consider the role of a ``minimal" setup with gray contours in Fig.~\ref{fig:mnu_As_contours}, which corresponds to removing DESI BAO and the $\tau$ prior from the baseline setup, i.e. corresponds strictly to unlensed $T$ and $E$ spectra with CMB S4. This minimal setup yields $\sigma (\sum m_\nu) = 0.359 \, \rm{eV}$. We also consider the combination of this minimal CMB S4 with galaxy tomography (green dashed) and galaxy and kSZ tomography (orange dashed), which yield $\sigma (\sum m_\nu) = 0.0714 \, \rm{eV}$ and $0.0676\, \rm{eV}$ respectively. The crucial point here is that the fractional improvement in $\sigma(\sum m_\nu)$ from kSZ tomography over the galaxy power spectrum alone (comparing solid orange and green curves), which was about $10\%$, is already significantly reduced, down to $\approx \!5\%$, once the information from the minimal CMB setup has been added. In other words, the galaxy power spectrum and the CMB experiment used for the kSZ velocity reconstruction already contain sufficient information about massive neutrinos to render the information in the reconstructed velocity field to be partially redundant.

We also note that while the overall constraining power and, in principle, the relative improvement from kSZ tomography are sensitive to the cosmological model, we found minimal changes in the latter when adding either curvature, $\Omega_k$, or extra relativistic species, parameterized in terms of an effective number of massless neutrinos $N_{\rm{eff}}$.

In the remaining subsections, we briefly analyze the extent to which these conclusions are sensitive to various modeling or survey assumptions.

\subsection{Bias Model \label{subsubsec:biases}}
Our baseline results have used the galaxy bias model defined in Eq.~(\ref{eq:gal_bias}), including linear bias, the RSD bias, and the gradient bias, as well as the velocity reconstruction bias defined in Eq.~(\ref{eq:bv_def}) that accounts for mismodeling of the galaxy-electron cross-correlation on small scales. Here we examine how our results depend on the details of the bias model. To this end, we repeat our analysis for a less conservative bias model with two changes: first, we adopt a linear-fit model for the linear bias as a function of redshift, $b_1(z) = mz + b_0$, and second, we take the RSD, gradient, and velocity reconstruction biases to be redshift independent, so that each is described by a single parameter. This significantly reduces the number of bias parameters in the redshift-summed Fisher matrix. Comparing the two bias models allows us to gauge the importance of the degeneracy between the bias parameters and the redshift-dependent effects that massive neutrinos imprint on the matter power spectrum and growth rate.

\begin{figure}
    \centering
\includegraphics[width=1.\columnwidth]{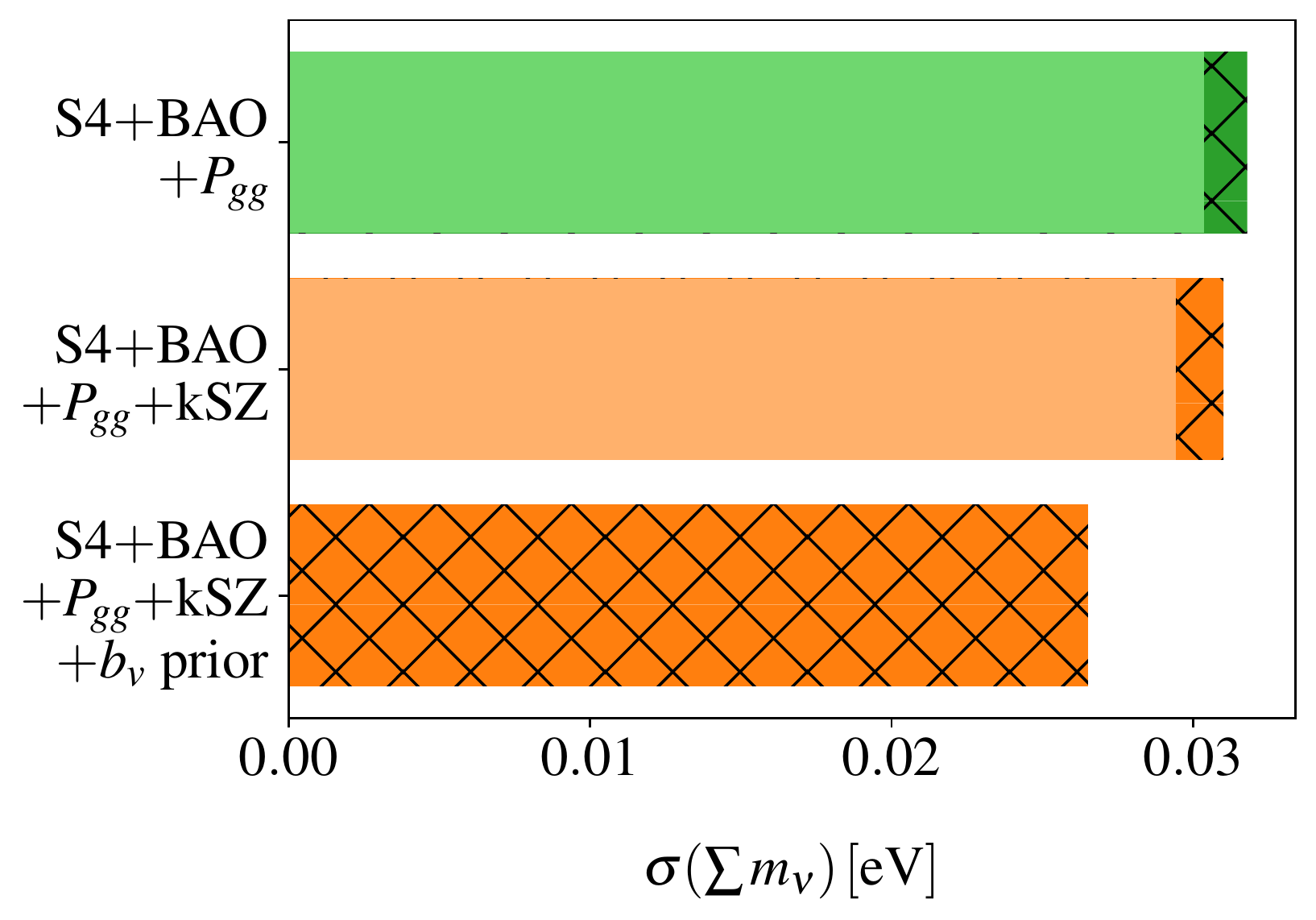}
    \caption{Effect of bias modeling on multi-tracer neutrino mass constraints. The upper two bars show the impact of galaxy bias modeling:  lighter, solid bars utilize a less conservative galaxy bias model as described in the body of the main text, whereas the darker, cross-hatched bars correspond to our baseline forecasts. The bottom orange bar also corresponds to the baseline forecast, but with the addition of a $1\%$ prior on the kSZ velocity reconstruction bias $b_v(z)$. The same $\tau$ prior as the baseline results, $\sigma(\tau) = 0.0075$, is used for all three bars.}
    \label{fig:mnu_gal_bias}
\end{figure}

The upper two bars in Figure~\ref{fig:mnu_gal_bias} show the impact that the less conservative bias model has on neutrino mass constraints from the CMB, BAO, and galaxy and kSZ tomography, with all other survey and model assumptions matching those of our baseline forecast. We find the constraints on $\sum m_\nu$ improve by $\approx \! 5\%$ when using the simpler bias model, and the improvement granted from kSZ tomography is still small, $\approx \! 3\%$. The minimal change in the constraints indicates that, with the background cosmology significantly pinned down by BAO and the CMB, the additional neutrino mass information coming from tracers of the late-time matter power spectrum---i.e., predominantly the galaxies---is not strongly limited by degeneracy with redshift-dependence in the astrophysical bias parameters. 

We further tested the impact of bias uncertainty by fixing one set of the galaxy bias parameters (e.g. fixing all $b_1(z)$, or all $b_{2}(z)$), as well as by imposing a $1\%$ prior on the kSZ velocity reconstruction bias $b_v(z)$. We emphasize that this prior, while optimistic, is achievable with various techniques, for example measurement of fast radio burst dispersion \cite{Madhavacheril:2019buy}. Among these tests, we find that in particular the $b_v$ prior, i.e., mitigating the kSZ optical depth degeneracy, can dramatically improve the efficacy of kSZ tomography: the measurement error with this prior is $\sigma(\sum m_\nu) =0.0265\, \rm{meV}$, corresponding to a nearly $17\%$ reduction compared to S4$+$DESI BAO+$P_{gg}$. A strong prior on $b_v$ grants kSZ tomography much greater sensitivity to amplitude effects, namely here the small, roughly flat suppression of large-scale matter power from massive neutrinos. With the $b_v$ prior, we find that even when ignoring scale-dependent growth information, adding kSZ tomography still provides a nearly $14\%$ reduction in $\sigma(\sum m_\nu)$, indicating that the gains are not simply from the imprint of neutrinos on $f(k)$. Here, the $\tau$ prior sufficiently tightens the $A_s$ constraint from the CMB, such that with the kSZ optical depth degeneracy mitigated, kSZ tomography can better distinguish the primordial amplitude from suppression effects from massive neutrinos. Note that the improvement vanishes if instead $b_v$ is marginalized in the forecast but assumed to be redshift-independent, which indicates that it is indeed pure amplitude degeneracies, rather than redshift-dependent information, that is important. By contrast, fixing terms in the galaxy bias has little effect on the improvement granted by kSZ tomography; even though $P_{gv}$ is sensitive to these, the kSZ optical depth degeneracy remains, and furthermore the constraining power of $P_{gg}$ is also improved by simplifying the galaxy bias model. In short, in a multi-tracer context, imperfect knowledge of $b_v$ presents a significant obstacle to neutrino mass inference with kSZ tomography, but the fine details of galaxy bias modeling do not.

We note that a caveat to this second conclusion would potentially exist if there was a signal of the neutrino mass in the galaxy bias itself, for example, as has been detected in N-body simulations \cite{LoVerde:2014pxa, Chiang:2017vuk, Chiang:2018laa}, a possibility that we discuss further in Sec.~\ref{sec:discussion}.

\subsection{CMB Lensing \label{sec:lensing}}

\begin{figure}[t!]
    \centering
\includegraphics[width=1.\columnwidth]{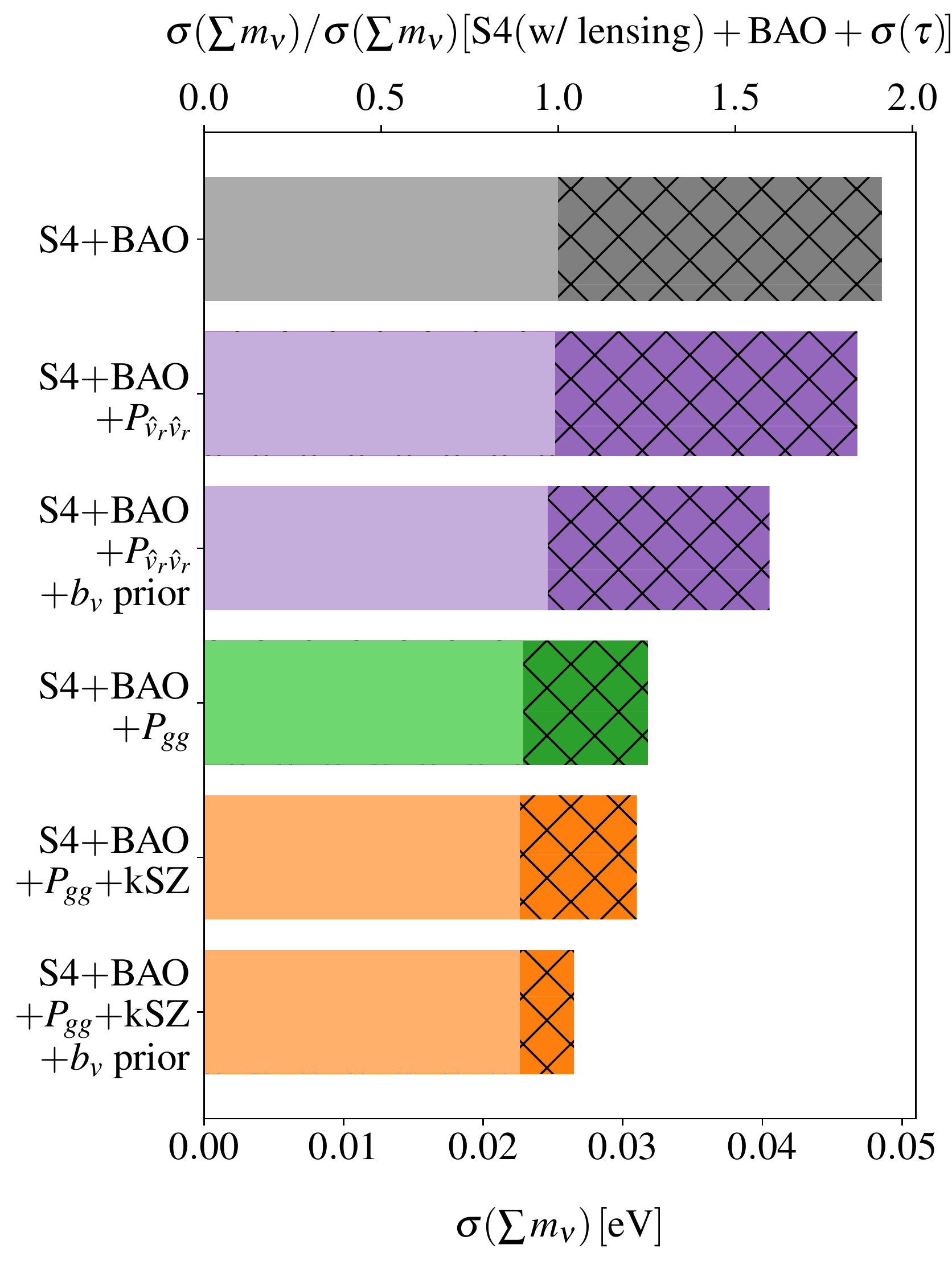}
    \caption{Impact of CMB lensing information on multi-tracer neutrino mass constraints. Darker, cross-hatched bars correspond to our baseline forecasts, which exclude CMB lensing, while lighter, solid bars include CMB lensing, as described in the body of the main text. We also include results when imposing a $1\%$ prior on $b_v(z)$, as in the previous subsection. All results shown here include a Planck-like $\tau$ prior. The lower horizontal axis denotes the neutrino mass constraint in $\rm{eV}$ while the upper horizontal axis shows the relative change in this constraint compared to the constraint from CMB S4 with lensing, DESI BAO, and a Planck-like $\tau$ prior.}
    \label{fig:mnu_lensing}
\end{figure}

Our baseline forecasts in Sec.~\ref{sec:forecast_baseline} have excluded all CMB lensing information, focusing on the role of kSZ tomography as an independent probe of the growth of large-scale structure. Those results indicate that kSZ tomography does not provide significant information about massive neutrinos in addition to the galaxy survey and the CMB (unless the kSZ optical depth degeneracy can be overcome). However, CMB lensing provides another probe of structure growth, the role of which we briefly examine here. CMB lensing, like galaxy (and in principle, kSZ) tomography, is a powerful measurement of late-time density fluctuations (and the growth thereof), providing a crucial probe of the scale- and redshift-dependent suppression of structure characteristic of massive neutrinos.
To examine the information in CMB lensing, we repeat our baseline forecast, but use lensed temperature and polarization spectra, and add the lensing reconstruction power spectrum to the observables. In Fig.~\ref{fig:mnu_lensing}, we compare the results of the setup with lensing information (lighter, solid bars) to the baseline forecast (darker, cross-hatched bars), finding that lensing information considerably improves constraints on massive neutrinos, reducing measurement error by approximately $25$-$50\%$ depending on the set of observables. The reference configuration of CMB S4, DESI BAO, with the Planck-like $\tau$ prior, yields $\sigma(\sum m_\nu) =0.0254\,\rm{eV}$. The LSST galaxy power spectrum is less impactful than in the baseline configuration, but still provides a $\approx 10\%$ reduction in $\sigma(\sum m_\nu)$, yielding $0.0229\,\rm{eV}$. This would enable nearly $3\sigma$ evidence for the normal hierarchy fiducial value of $\sum m_\nu =0.06 \, \rm{eV}.$ The additional constraining power from kSZ tomography is $\approx 1\%$, even more limited than it was in the baseline setup, due to the less singular role of kSZ when lensing provides (significant) information about the growth of structure. 

It is important to note that the optical depth prior becomes important for neutrino mass constraints when CMB lensing information is included. CMB S4 will only probe multipoles $\ell \gtrsim 30$, and hence cannot measure the large-scale $E$-mode reionization bump\footnote{Beyond measurement of the reionizaiton bump in the CMB, 21cm measurements offer a promising route to independently reduce the uncertainty on $\tau$ and further improve neutrino mass constraints \cite{Liu:2015txa,Shmueli:2023box}}, inhibiting the breaking of the $A_s$-$\tau$ degeneracy. 
With lensing information included, the Planck-like $\tau$ prior reduces the measurement uncertainty by almost one-half for the reference configuration (S4+DESI BAO), and by approximately a quarter for configurations including the LSST galaxy power spectrum, by breaking the overall amplitude degeneracy between $A_s$, $\tau$, and $\sum m_\nu$ in the CMB spectra. 

We also show in Fig.~\ref{fig:mnu_lensing} neutrino mass constraints both with and without lensing when the $1\%$ $b_v(z)$ prior is included. These results show that with the inclusion of CMB lensing information, the impact of the $b_v(z)$ prior becomes negligible. We find this is still true if we use lensed $T$ and $E$ spectra but exclude the information from the lensing reconstruction power spectrum. The neutrino mass information from CMB lensing is significant enough to overshadow any contribution from kSZ, even when significantly mitigating the kSZ optical depth degeneracy. This is expected: CMB lensing is highly sensitive to the growth of cosmic structure and is a direct, unbiased probe of the total matter density fluctuations, leaving even nearly-unbiased velocities with a less distinguishing role. Broadly, these results reinforce the growth of large-scale structure as a powerful source of information about massive neutrinos for stage IV surveys. CMB lensing fills this role definitively, while kSZ tomography, although less powerful, can do so complementarily, as a sole growth probe, if its optical depth degeneracy can be surmounted.

\subsection{Planck}
\begin{figure}[b!]
    \centering
\includegraphics[width=1.\columnwidth]{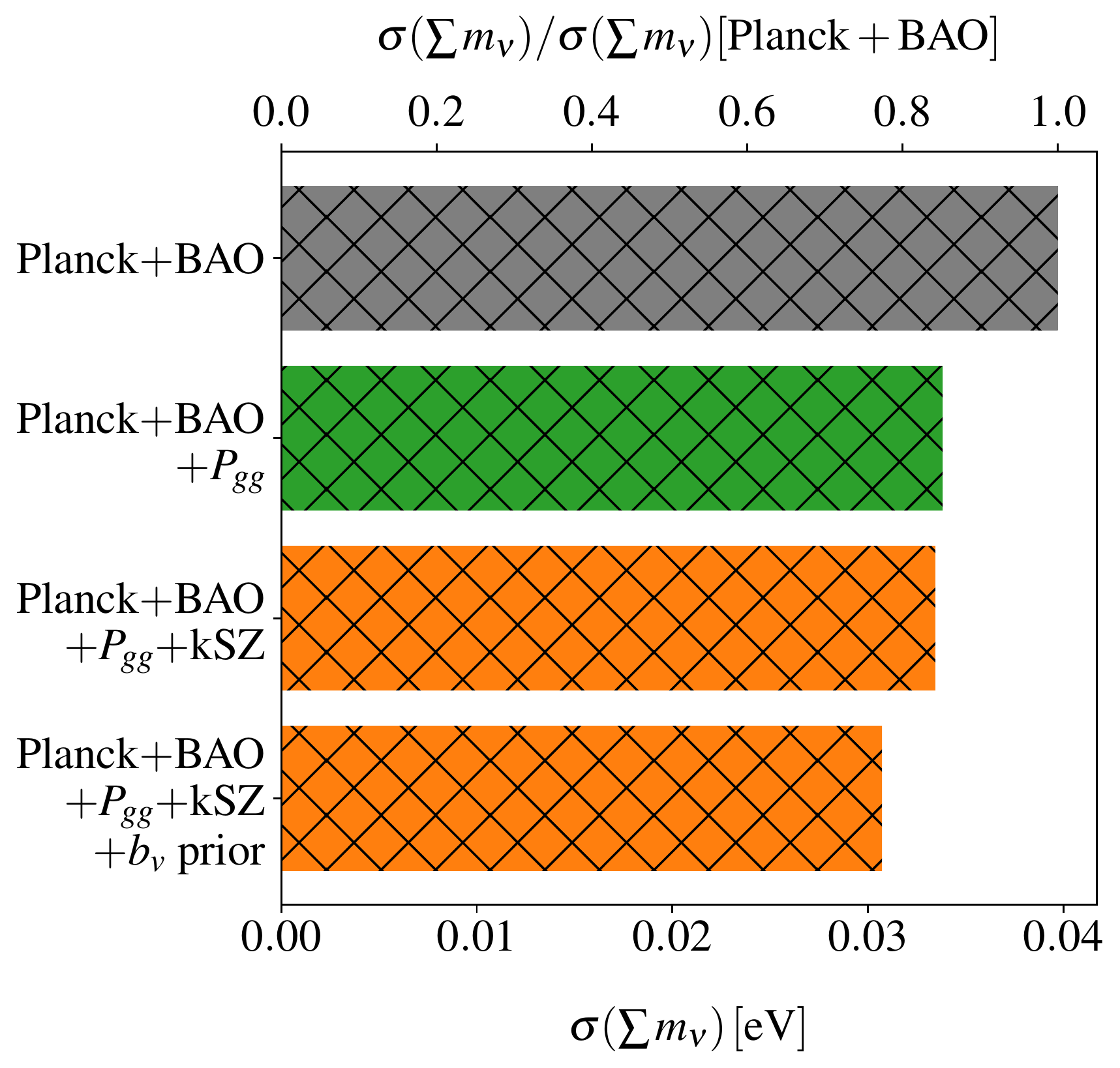}
    \caption{Multi-tracer neutrino mass constraints with Planck PR3 and BAO from Ref.~\cite{Aghanim:2018eyx}, without the Planck lensing reconstruction spectrum. The galaxy and kSZ tomography Fisher matrices still follow the baseline forecast specifications, that is they are assumed to have specifications corresponding to LSST and CMB S4. We also show results when including a $1\%$ $b_v(z)$ prior. The upper horizontal axis shows the relative change in the neutrino mass constraint compared to the constraint from Planck + BAO (gray).  }
    \label{fig:mnu_Planck}
\end{figure}
We have assumed in our baseline forecast that both the kSZ velocity reconstruction and information from the CMB temperature and polarization spectra are furnished by an S4-like CMB experiment, focusing on the role of kSZ tomography in the presence of the contemporaneous CMB primary information. Here, we instead briefly consider whether kSZ velocity reconstruction with S4- and LSST-like surveys can improve neutrino mass inference when combining with previous generation CMB and BAO information. We examine such a scenario in Fig.~\ref{fig:mnu_Planck}, where we take the PR3+BAO data release~\cite{Aghanim:2018eyx}, without the lensing reconstruction spectrum. Note there is a key difference here compared to our baseline results, which take unlensed $T$ and $E$ CMB spectra, because the measured Planck spectra are lensed and hence contain additional information about massive neutrinos. Qualitatively, however, the results here are similar to our baseline results. The overall constraints are different (for example, the (lensed) PR3 Planck$+$BAO covariance yields a $0.04 \, \rm{eV}$ $1\sigma$ error bar, compared to $0.049 \, \rm {eV}$ from unlensed S4$+$DESI BAO), but the improvement from adding the LSST galaxy power spectrum to the Planck analysis is still valuable, providing an $\mathcal{O}(15\%)$ reduction in $\sigma (\sum m_\nu)$, and further including kSZ tomography yields a similarly small ($\mathcal{O}(1\%)$) additional reduction. Comparing with our baseline results, this indicates that PR3 Planck+BAO already provides sufficient information to render the kSZ-driven improvement over the galaxy power spectrum alone, seen in Fig.~\ref{fig:mnu_As_contours}, redundant, without moving to S4 or DESI level sensitivities. However, we find that kSZ can provide some improvement in neutrino mass constraints, nearly $10\%$, with a $1\%$ $b_v(z)$ prior. This is in contrast to the results from the previous subsection, which showed that with Stage IV surveys, the information from the lensing of the temperature and polarization spectra was sufficient to overshadow the effect of a $b_v$ prior.  
Altogether, this suggests that Stage IV kSZ tomography will remain only marginally useful when applied to neutrino mass analyses with Planck-era data, although some gains can be made in the special case of a strong $b_v(z)$ prior.

\subsection{Futuristic Scenario \label{sec:futuristic}}
\begin{figure}[t!]
    \centering
\includegraphics[width=1.\columnwidth]{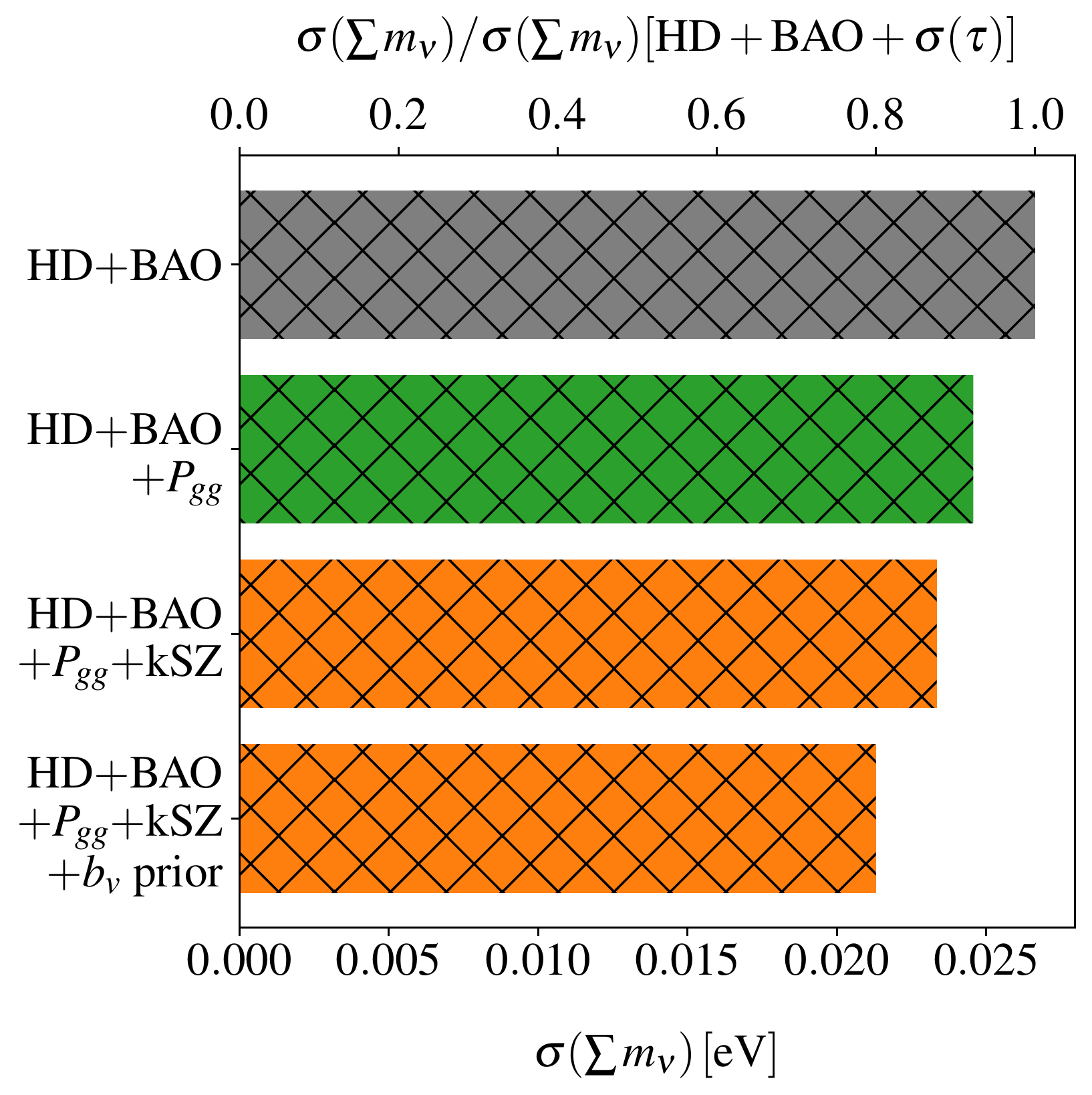}
    \caption{Multi-tracer neutrino mass constraints with an HD-like CMB experiment, with no lensing information. To isolate the effect of the higher resolution CMB experiment, baseline forecast assumptions are maintained for BAO (DESI) and the galaxy survey (LSST). The relative change in the constraint, compared to the HD$+$BAO$+\sigma(\tau)$ setup, is indicated via the upper horizontal axis.}
    \label{fig:mnu_HD}
\end{figure}

We extend our analysis by considering a futuristic CMB-HD-like experiment \cite{Sehgal:2019ewc}, where we assume $\theta_{\rm{FWHM}} = 20''$ and $\Delta_T = 0.1\, \mu K '$. Although a futuristic CMB experiment would possibly be accompanied by a contemporaneous galaxy survey and/or BAO measurement, e.g. MegaMapper \cite{Schlegel:2019eqc}, our point here is to isolate the specific implications that a higher-resolution CMB experiment has for constraining $\sum m_\nu$ with kSZ tomography, and hence we hold other details of the forecast fixed (i.e. DESI BAO, LSST $P_{gg}$, and no lensing information). The results of this setup are shown in Fig.~\ref{fig:mnu_HD}. We find that with HD + BAO,  $\sigma(\sum m_{\nu})$ is reduced significantly compared to the setup with S4, reaching $\sigma (\sum m_\nu) \approx 0.0266 \,\rm{eV}$. Adding the LSST galaxy power spectrum information, the measurement error reaches $\sigma(\sum m_\nu) = 0.0246 \, \rm{eV}$, and further adding kSZ tomography brings this to $\sigma(\sum m_\nu) = 0.0233 \, \rm{eV}$. That is, with a significantly higher resolution CMB experiment, the reduction in measurement error in $\sum m_\nu$ from the inclusion of kSZ tomography is only $\approx \! 5\%$. It has been previously shown that with LSST, the kSZ velocity reconstruction noise experiences diminishing returns for improved beam size and white noise levels much beyond those of roughly S4 levels \cite{Munchmeyer:2018eey}, which we also see reflected here in the marginal gains by pushing to HD-like specifications. As before, we also consider the effect of a $1\%$ prior on $b_v(z)$, which yields $\sigma(\sum m_\nu) = 0.0213 \, \rm{eV}$, an approximately $13\%$ reduction. These results largely echo those of our baseline forecast, indicating the role of kSZ tomography as an independent growth probe, in the context of neutrino mass constraints, significantly depends on the breaking of the kSZ optical depth degeneracy.

Finally, we consider a futuristic setting that pairs the HD-like CMB experiment with a futuristic galaxy survey with LSST-like number densities, but spectroscopic redshifts. For simplicity, we do this by keeping the LSST survey specifications the same, but setting $\sigma_z = 0$ in Eq.~(\ref{eq:Wphotoz}). This improves the constraining power of not only the galaxy power spectrum, but also of kSZ tomography, which benefits greatly from improved reconstruction of radial modes. In this case, we find the combination of CMB HD (without lensing), DESI BAO, a prior $\sigma(\tau) = 0.0075$, and the futuristic galaxy survey gives $\sigma(\sum m_\nu) = 0.0219\, \rm{eV}$. The addition of kSZ tomography would yield an approximately $25\%$ tighter constraint, $\sigma(\sum m_\nu) = 0.0165\, \rm{eV}$.  A contemporary BAO measurement and optical depth prior would also likely accompany the setup that we have explored here, but nevertheless, these results suggest that in a futuristic setting, kSZ tomography may offer a powerful growth probe for the purpose of neutrino mass constraints.

\section{Discussion}\label{sec:discussion}
In this paper we have examined the role of kSZ tomography as an independent probe of growth for cosmological neutrino mass constraints with Stage IV CMB and LSS surveys. Our primary result, summarized by Figs.~\ref{fig:mnu_baseline} and \ref{fig:mnu_As_contours}, is that kSZ tomography does provide information about massive neutrinos, predominantly via their effect on scale-dependent growth, but that this information is eclipsed by the constraining power of the CMB and galaxy power spectrum. 

We investigated the extent to which the marginal improvement from kSZ tomography depends on modeling and survey assumptions, including galaxy bias,  the kSZ optical depth degeneracy, cosmological models, and CMB and BAO information.  Our results consistently show that the inclusion of primary CMB temperature and $E$-mode polarization anisotropies nearly eliminates kSZ-driven improvement in neutrino mass constraints, even without lensing information, BAO, or a strong measurement of $\tau$. As outlined in Sec.~\ref{sec:intro}, there are several distinct avenues by which kSZ tomography could, in principle, improve neutrino mass inference, however in practice we find these all to offer only marginal additional constraining power. As our baseline forecasts demonstrate, kSZ as the sole probe of the growth of cosmic structure can tighten neutrino mass constraints, specifically via measurement of scale-dependent effects in the growth rate, $f(k)$, but this contribution is only significant compared to the galaxy power spectrum in isolation, and becomes subdominant in a multi-tracer context. On the other hand, our analysis shows that if the kSZ optical depth degeneracy can be overcome, the additional amplitude information from kSZ can appreciably improve neutrino mass constraints. However, this only holds true in the absence of all CMB lensing information, in which case kSZ has a distinct role as the singular (nearly) unbiased tracer of density fluctuations. The information from the lensing of CMB temperature and polarization spectra is sufficient to almost entirely diminish this effect, even when neglecting information from the lensing reconstruction spectrum. This also suggests that, in a multi-tracer setting, the role of kSZ-reconstructed velocities as an anchor for the galaxy power spectrum only meaningfully materializes if the velocities are effectively the sole unbiased tracer of the late time density field, which is not the case in the presence of lensing. Finally, our tests with various modifications to the galaxy bias model showed minimal impact, demonstrating that sample variance cancellation to break astrophysical and cosmological parameter degeneracies provides limited additional neutrino mass information. Ultimately, the information added by kSZ tomography is largely redundant, as it is already captured by the combination of CMB and galaxy surveys used for velocity reconstruction.

The above applies to the imprint of massive neutrinos on the matter power spectrum and growth rate, but an additional consideration involves the scale-dependent galaxy bias induced by massive neutrinos, which presents a potential independent signal for constraining neutrino mass~\cite{LoVerde:2014pxa, Chiang:2017vuk}. This is a scenario where kSZ tomography is particularly advantageous, as it provides an independent probe of the matter power spectrum with high signal-to-noise on large scales, primarily via the galaxy-velocity cross-spectrum. Using differently biased tracers of the matter power spectrum, kSZ tomography enables the estimation of bias ratios while canceling large-scale sample variance, facilitating the detection of physical effects embedded in the large-scale bias. For massive neutrinos, this manifests as a step-like feature sensitive to the total neutrino mass~\cite{Chiang:2018laa}. Although the signal for the normal hierarchy is relatively weak, kSZ tomography may enhance sensitivity to this scale-dependent bias, improving constraints on neutrino mass. Furthermore, if this feature is inadequately measured or modeled, it could introduce biases or degeneracies in parameter inference, which kSZ tomography may mitigate more effectively than the galaxy power spectrum alone. These implications require further dedicated analysis, to be addressed in a forthcoming study~\cite{Tishue:inprep}. 

While the role of kSZ tomography for neutrino mass constraints may remain limited in Stage IV surveys, our results nevertheless demonstrate that growth information, namely from CMB lensing, will remain important. Were kSZ to provide a probe of growth competitive with lensing, for example in the futuristic scenario examined in Sec.~\ref{sec:futuristic}, it may play an important role in a precise cosmological determination of the total neutrino mass. 

Looking forward, future experiments will continue to refine neutrino mass measurements through complementary approaches, such as precise optical depth measurements, BAO and full-shape analyses, and innovative techniques like line intensity mapping~\cite{2019BAAS...51g..58K, Booth:2024jkg, Booth:2024oem, Shmueli:2024npx}, as well as searches for new signatures~\cite[e.g][]{Nascimento:2023ezc}. These advancements, together with novel applications of joint CMB and large-scale structure analyses, particularly in disentangling scale-dependent biases induced by neutrinos, could uncover new avenues to improve the robustness and precision of cosmological bounds on the total neutrino mass.

\section{Acknowledgments}
We thank Robert Caldwell, Neal Dalal, Gil Holder, Marc Kamionkowski, Sarah Libanore, Marilena Loverde, Jos\'e~Luis~Bernal, Srinivasan Raghunathan,  Gabi~Sato-Polito, Gali Shmueli, Kendrick Smith, and Cynthia Trendafilova for useful conversations and suggestions. AJT is supported by the United States Department of Energy, DE-SC0015655. AJT gratefully acknowledges the support of the Centre for Cosmological Studies Balzan Fellowship, during which this work was initiated. SCH is supported by P.~J.~E.~Peebles Fellowship at the Perimeter Institute for Theoretical Physics. This research was supported in part by Perimeter Institute for Theoretical Physics. Research at Perimeter Institute is supported by the Government of Canada through the Department of Innovation, Science and Economic Development Canada and by the Province of Ontario through the Ministry of Colleges and Universities. EDK acknowledges joint support from the U.S.-Israel Bi-national Science Foundation (BSF, grant No.\ 2022743) and the U.S.\ National Science Foundation (NSF, grant No.\ 2307354), and support from the ISF-NSFC joint research program (grant No.\ 3156/23). MM acknowledges support from NSF grants AST-2307727 and  AST-2153201 and NASA grant 21-ATP21-0145.  This work was also carried out in part at the Advanced Research Computing at Hopkins (ARCH) core facility  (rockfish.jhu.edu), which is supported by the National Science Foundation (NSF) grant number OAC1920103. This work was performed in part at Aspen Center for Physics, which is supported by National Science Foundation grant PHY-2210452.

\bibliography{main}

\end{document}